\documentclass[singlecolumn,showpacs,preprintnumbers,amsmath,amssymb]{revtex4}

\usepackage{graphicx}
\usepackage{dcolumn}
\usepackage{bm}

\begin{document}

\title{RNA matrix models with
external interactions and their asymptotic behaviour}

\author{I. Garg}
\author{N. Deo}
 \email{ndeo@physics.du.ac.in}
\affiliation{Department of Physics \& Astrophysics\\ University
of Delhi, Delhi 110007, India}

\begin{abstract}
We study a matrix model of RNA in which an external perturbation
acts on n nucleotides of the polymer chain. The effect of the
perturbation appears in the exponential generating function of the
partition function as a factor $(1-\frac{n\alpha}{L})$ [where
$\alpha$ is the ratio of strengths of the original to the perturbed
term and L is length of the chain]. The asymptotic behaviour of the
genus distribution functions for the extended matrix model are
analyzed numerically when (i) $n=L$ and (ii) $n=1$. In these matrix
models of RNA, as $n\alpha/L$ is increased from 0 to 1, it is found
that the universality of the number of diagrams $a_{L, g}$ at a
fixed length L and genus g changes from $3^{L}$ to
$(3-\frac{n\alpha}{L})^{L}$ ($2^{L}$ when $n\alpha/L=1$) and the
asymptotic expression of the total number of diagrams $\cal N$ at a
fixed length L but independent of genus g, changes in the factor
$\exp^{\sqrt{L}}$ to $\exp^{(1-\frac{n\alpha}{L})\sqrt{L}}$
($exp^{0}=1$ when $n\alpha/L=1$).
\end{abstract}

\maketitle

\section{\label{sec:level1}INTRODUCTION}
Improved understanding of the process of folding of RNA finds its
ultimate use in the prediction of the fully folded, partially folded
and completely unfolded structures under physiological conditions
\cite{1}. Under these conditions, unfolding is a very slow process
as compared to folding in the presence of a force. Application of a
force increases the unfolding rate and we can therefore get the
unfolded structures from the folded ones (\cite{1} and references
therein). Experimental techniques of force induced measurements have
proved successful in probing properties related to different aspects
of RNA folding and unfolding, domain unfolding in proteins, in
polysachharides and nucleic acids (\cite{2} and references therein).
Experiments have been performed on the double helixed DNAs to study
their elastic and structural properties using electric field,
hydrodynamic flow among other methods of force application (\cite{3}
and references therein). The advent of AFM technique served as an
important tool in the study of the basic underlying framework of
molecular structural biology. Over the years, optical tweezers and
AFM (atomic force microscopy) techniques have been employed to study
the physical, elastic and structural properties of the biomolecules
by recording their force extension curves (FECs) and studying the
force dependent dynamics and folding landscapes of the molecules
(\cite{4,5,6,7,8,9} and references therein). The conformations of
biopolymers (DNA, RNA and proteins) which are otherwise not
accesible from the conventional methods of measurements: NMR
spectroscopy and X-ray crystallography, are possible with the use of
AFMs. These conformations help in revealing the underlying
mechanical framework of the biological systems (\cite{2} and
references therein). Mechanical unfolding and refolding of single
RNA has been studied using force-ramp, hopping and force-jump
methods (\cite{10} and references therein). In mechanical unfolding
experiments, it has been observed that at a critical value of the
applied force, the hairpin structure toggles between the folded and
the unfolded states \cite{11,12,13}. In these experiments, ionic
concentrations play an important role. Experiments of Bustamante
\textit{et al} \cite{12,13} have shown that the denaturation of RNA
by a constant force involves multiple trajectories (for RNA hairpins
and Tetrahymena thermophila ribozyme) while undergoing a transition
from the folded structure state to the unfolded state. These
trajectories depend on the point at which the force is applied
\cite{1,14}. This diverseness in the folding-unfolding pathways is
due to the rugged energy landscape of RNA (consisting of many
minima). Controlled/monitored force loading and unloading rates can
be used to manipulate the single molecules of RNA into either their
native or misfolded pathways. Different force unloading rates in
experiments on TAR RNA molecules showed different types of
trajectories associated with particular refolding characteristics
(\cite{15} and references therein).

We discuss here very briefly, a generalization of the extended
random matrix model of RNA folding proposed in \cite{16} where the
external perturbation acts on a single nucleotide ($n=1$) and on n
nucleotides ($n\leq L$) in the polymer chain (we will refer to the
two RNA models as 1-NP RNA model, with NP being Nucleotide
Perturbation and n-NP RNA model respectively). In \cite{16}, the
external perturbation acted on all the nucleotides in the polymer
chain (\textit{i.e.,} $n=L$, n is the number of bases on which the
force is acting). We briskly outline the extended matrix model of
\cite{16} for completeness and understanding and follow it up with
results and comparative discussion for the 1-NP and n-NP models.
Further, we present a detailed numerical analysis for the
asymptotics of the extended matrix model of RNA with perturbation on
all the nucleotides in the polymer chain. The genus distribution
functions: the total number of diagrams at a fixed length L but
independent of genus g, $\cal N$ and the number of diagrams at a
fixed length L and genus g, $a_{L, g}$ of the matrix model of RNA in
\cite{17,18} are found to change in the presence of the external
perturbation. We extend our numerical asymptotic analysis to the
n-NP RNA model as well.

\section{EXTENDED MATRIX MODELS OF RNA}

We review here, the effect when a perturbation acts on all the
nucleotides in the polymer chain ($n=L$) studied in \cite{16}. The
nucleotide-nucleotide interaction partition function of the polymer
chain with a perturbation on all the bases is

\begin{equation}
\label{eq.1} Z_{L, \alpha}(N) = \frac{1}{A_{L}(N)} \int
\prod_{i=1}^L d\phi_{i} exp^{-\frac{N}{2} \sum_{i, j = 1}^L
(V^{-1})_{i, j} Tr \phi_{i} \phi_{j}} exp^{-N \sum_{i=1}^{L}
(W^{-1})_{i} Tr \phi_{i}} \frac{1}{N} Tr \prod_{i = 1}^L (1 +
\phi_{i})
\end{equation}

where $A_{L}(N)=\int \prod_{i = 1}^{L} d\phi_{i} exp^{-\frac{N}{2}
\sum_{i, j = 1}^L (V^{-1})_{i, j} Tr \phi_{i} \phi_{j}}exp^{(-N)
\sum_{i = 1}^L (W^{-1})_{i} Tr \phi_{i}}$ is the normalization
constant, $exp^{-N \sum_{i=1}^{L} (W^{-1})_{i} Tr \phi_{i}}$ is the
perturbation term, $V_{i, j}$ is an (L$\times$L) symmetric matrix
containing information on the interactions between the L nucleotides
at positions i and j in the polymer chain, $\phi_{i}$ are L
independent (N$\times$N) hermitian matrices and the observable
$\prod_{i} (1+\phi_{i})$ is an ordered product over $\phi_{i}$'s. We
consider $V_{i, j}=v$ and $W_{i}=w$ where v gives the strength of
interaction between the nucleotides at positions i and j (in these
models, interaction between any two nucleotides of the chain is
considered the same and equal to v) and w gives the strength of the
perturbation. Carrying out a series of Hubbard Stratonovich
Transformations, the integral over L matrices $\phi_{i}$ in
eq.~(\ref{eq.1}) reduces to an integral over a single (N$\times$N)
hermitian matrix $\sigma$

\begin{equation}
\label{eq.2} Z_{L, \alpha}(N) = \frac{1}{R_{L}(N)} \int d\sigma
exp^{-\frac{N}{2v} Tr (\frac{v}{w} + \sigma)^2} \frac{1}{N} Tr (1
+ \sigma)^L
\end{equation}

where $R_{L}(N)=\int d\sigma exp^{-\frac{N}{2v} Tr (\frac{v}{w} +
\sigma)^2}$. Following the algebra in \cite{16} (from eq.5 to
eq.15), the exponential generating function $G(t, N, \alpha)$ of the
partition function $Z_{L, \alpha}(N)$ is

\begin{equation}
\label{eq.3} G(t, N, \alpha) \equiv \sum_{L = 0}^{\infty}Z_{L,
\alpha}(N) \frac{t^L}{L!} = exp^{\frac{v t^2}{2N} + t(1 - \alpha)}
\left[ \frac{1}{N} \sum_{k = 0}^{N - 1} \binom{N}{k+1} \frac{(t^2
v)^k}{k! N^k} \right]
\end{equation}

where $\alpha=\frac{v}{w}$ gives the ratio of strengths of the
original to the perturbed term.

For $\alpha = 0$, the extended matrix model of RNA folding reduces
to the random matrix model in \cite{18}. However, for $\alpha = 1$
it is observed that the partition function for odd lengths of the
polymer chain vanishes completely. In the extended matrix model,
each unpaired base of the polymer chain in the contact diagrams is
associated with a factor $(1 - \alpha)$ which becomes zero when
$\alpha=1$ thus removing structures with any unpaired bases. We can
therefore divide the structures into two regimes: (i) $\alpha \leq
1$ comprising of both the unpaired and paired base structures and
(ii) $\alpha=1$ comprising of only completely paired base structures
(where only structures with fully paired bases remain whereas
structures with any unpaired bases are suppressed) \cite{16}. The
genus distributions for the extended matrix model are therefore
significantly different for different $\alpha$'s, especially for
$\alpha=1$ [where $Z_{L, \alpha}(N)=0$ for odd lengths of the
polymer chain] as compared with the model of \cite{18}. The addition
of a perturbation has thus changed the genus distributions and the
overall enumeration of the structures given by this model.

\subsection{EXTENDED MATRIX MODEL OF RNA WITH PERTURBATION ON A SINGLE BASE (1-NP) AND n BASES (n-NP)}

We now consider a generalization of the extended matrix model
proposed in \cite{16} by adding a perturbation to a single
nucleotide in the polymer chain only, $n=1$ (1-NP). The motivation
comes from the force induced experiments in obtaining important
characteristics of folding and unfolding of RNAs discussed in the
introduction \cite{1,2,3,4,5,6,7,8,9,10,11,12,13,14,15}. We keep all
the assumptions the same as for the model in \cite{16}. The
interaction partition function $Z_{L, \alpha}(N)$ for 1-NP will be
given by eq.~(\ref{eq.1}) with the perturbation term now being
$exp^{-N (W^{-1})_{1} Tr \phi_{1}}$ and the normalization constant
given by $A_{L}(N)$=$\int \prod_{i = 1}^{L} d\phi_{i}
exp^{-\frac{N}{2} \sum_{i, j = 1}^L (V^{-1})_{i, j} Tr \phi_{i}
\phi_{j}} exp^{(-N) (W^{-1})_{1} Tr \phi_{1}}$. Carrying out a
similar mathematical analysis employed in going from eq. (1) to eq.
(3) above we can write the exponential generating function of the
partition function as in eq. (3) with the only difference being that
$\alpha$ in eq. (3) gets replaced by $\frac{\alpha}{L}$ for the
1-NP. This implies that when $\frac{\alpha}{L}=0$ \textit{i.e.}, no
perturbation is acting, we get the matrix model of \cite{18}. When
$\frac{\alpha}{L}=1$, we get the extended matrix model with
perturbation on all the bases \cite{16}. The 1-NP partition
functions $Z_{L, \alpha}(N)$ for different L can be found exactly
from the exponential generating function [eq.~(\ref{eq.3}) with
$\alpha$ being replaced by $\frac{\alpha}{L}$] by equating the
coefficients of powers of t on both the sides of the equation. In
general, if the number of bases with the perturbation is n then
$\alpha$ is replaced by $\frac{n \alpha}{L}$. When $n=L$, we get the
extended matrix model with perturbation on all the bases, discussed
briefly here, [eq.~(\ref{eq.1})-eq.~(\ref{eq.3})] and in detail in
\cite{16}.

The diagrammatic representation of the n-NP differs from the
diagrammatics of the model with perturbation on all the nucleotides
in the factor $(1-\frac{n\alpha}{L})$ associated with each unpaired
base which replaces the factor $(1-\alpha)$ in the contact diagrams
of figure 1 in \cite{16}.

\section{ASYMPTOTICS OF THE EXTENDED MATRIX MODELS FROM NUMERICS}

The asymptotic behaviour of the genus distribution functions for the
matrix model of RNA studied in \cite{18} showed universal
characteristics. We investigate here numerically, the changes that
the genus distribution functions: (i) the total number of diagrams
at a fixed length L but independent of genus g, $\cal N$ [defined as
$\cal N$=$Z_{L}(N=1)$] and (ii) the number of diagrams at a fixed
length L and genus g, $a_{L, g}$ [defined through
$Z_{L}(N)$=$\sum_{g=0}^{\infty}a_{L, g}\frac{1}{N^{2g}}$] of the
model in \cite{18} undergo when a perturbation is added to these
models. The asymptotics of the genus distribution functions are
computed for the extended matrix model (i) with perturbation on all
the bases, $n=L$ \cite{16} and (ii) with perturbation on n bases,
(n-NP). We will represent the genus distribution functions for the
different matrix models as follows: (i) $\cal N$ and $a_{L, g}$ will
represent the asymptotic formulae for the model in \cite{18}, (ii)
${\cal N}^\prime_{\alpha}$ and $a^\prime_{L, g, \alpha}$ will
represent the new asymptotic formulae for the extended matrix model
of RNA \cite{16} and (iii) ${\cal N}_{\alpha}$ and $a_{L, g,
\alpha}$ will represent the numerical values of the genus
distribution functions for different $\alpha$'s. We start with the
exact asymptotic expressions (i) ${\cal N} = L^{\frac{L}{2}}
exp^{\left[-\frac{L}{2} + \sqrt{L} - \frac{1}{4}\right]}/\sqrt{2}$
and (ii) $a_{L, g} = k_{g} 3^{L} L^{(3g - \frac{3}{2})}$ from
\cite{18} and compare the behaviour of ${\cal N}_{\alpha}$ and
$a_{L, g, \alpha}$ for the extended matrix model for lengths upto
$L=40$ for different values of $\alpha(=0,0.25,0.5,0.75,1)$. We
begin by studying ${\cal N}_{\alpha}$ and $a_{L, g, \alpha}$ for the
extended models with perturbation on all the bases ($n=L$).

\begin{figure}
\includegraphics[width=8cm]{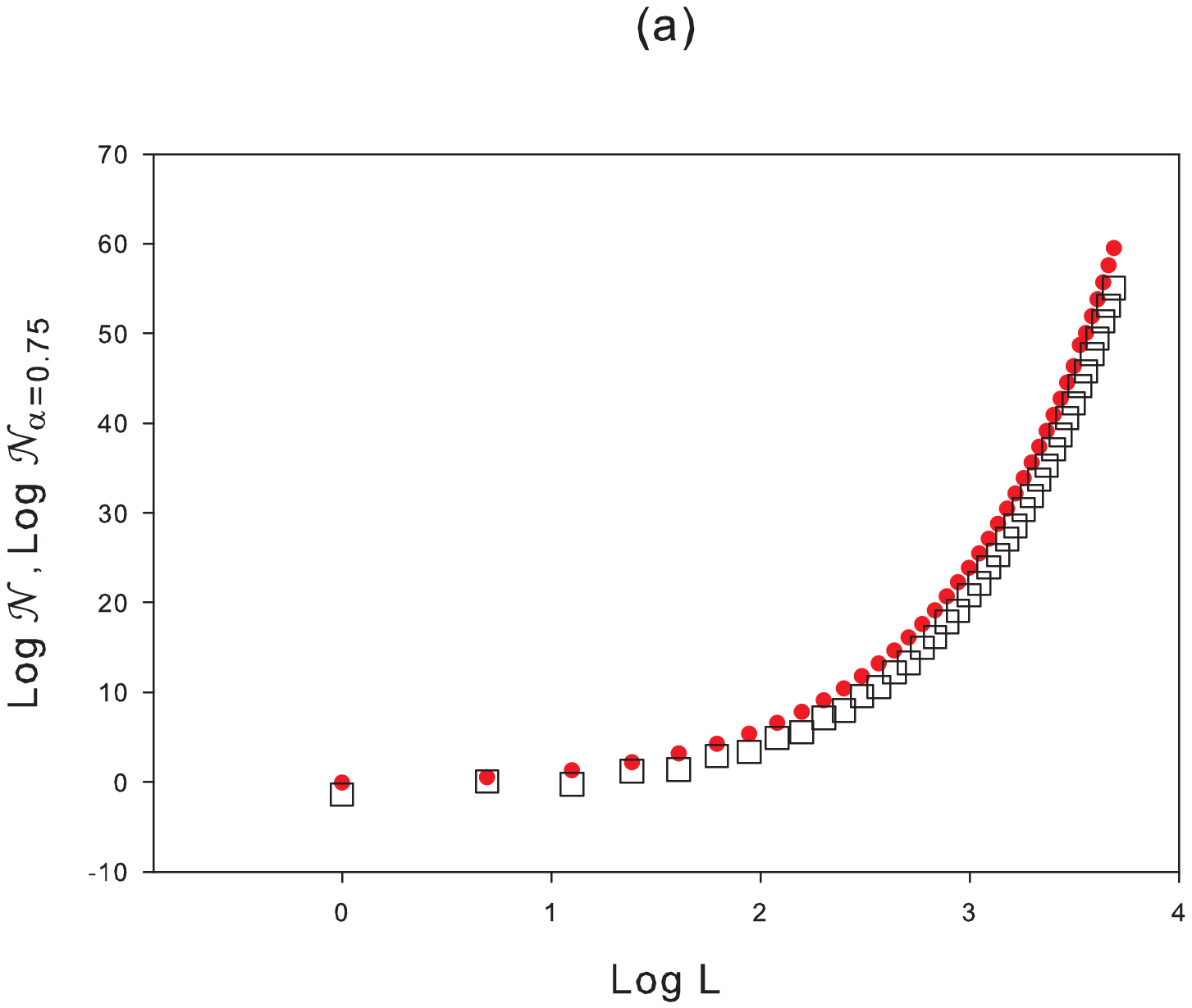}
\includegraphics[width=8cm]{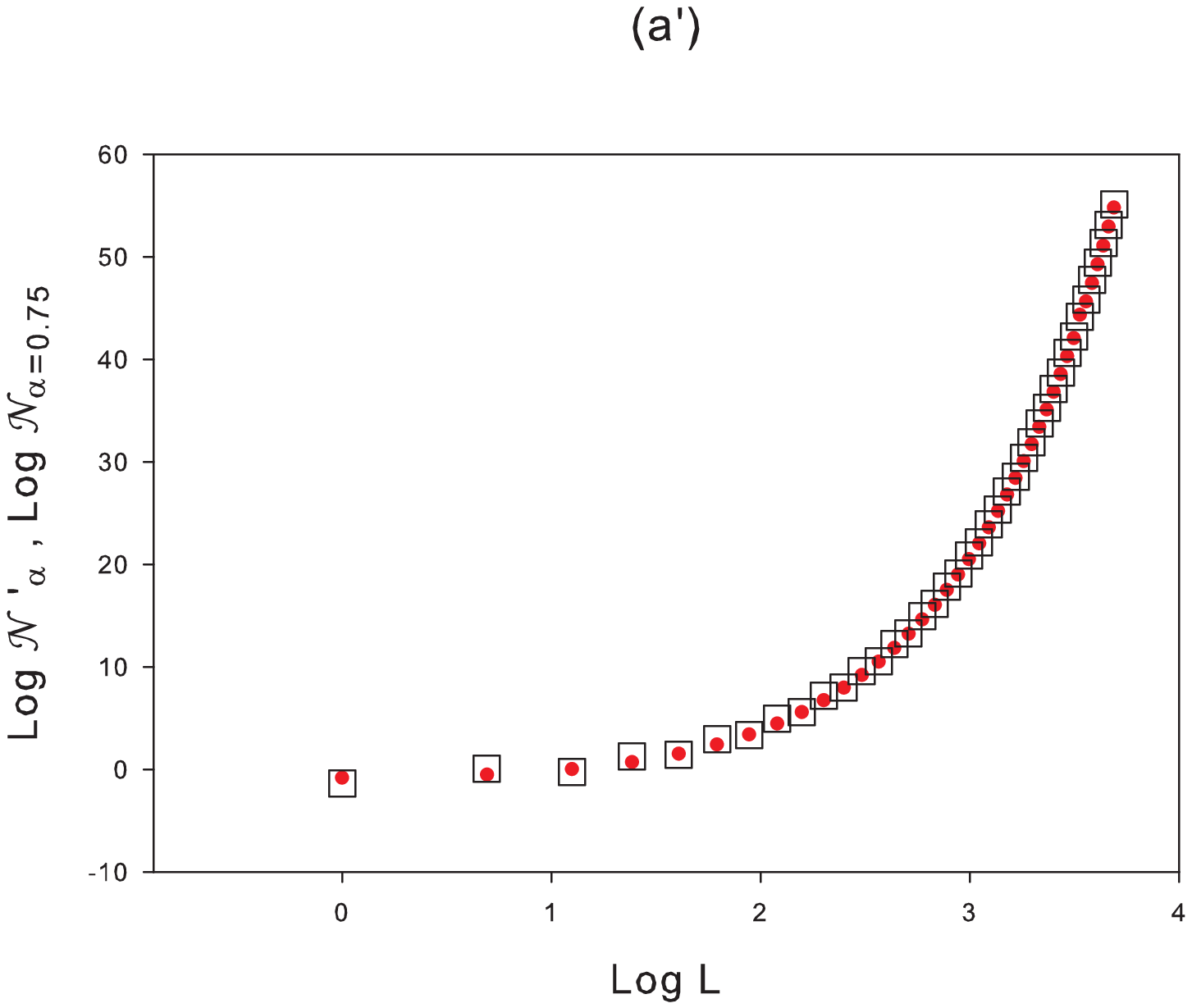}\\
\caption{(a) Plot of the asymptotic formula of $\cal N$ in \cite{18}
(red dotted curve) with the numerically calculated ${\cal
N}_{\alpha}$ values for different lengths L corresponding to
$\alpha=0.75$ (boxed curve).
\newline ($a^\prime$) The new asymptotic formula of ${\cal N}^\prime_{\alpha}$ (red dotted curve)
for the extended matrix model of RNA \cite{16} is plotted with the
numerical ${\cal N}_{\alpha}$ values for different lengths L for
$\alpha=0.75$ (boxed curve).} \label{fig.1}
\end{figure}

\subsubsection{Asymptotics for ${\cal N}_{\alpha}$}

Figure 1(a) shows the combined plot of the asymptotic expression
of $\cal N$ (red dotted curve) with the numerically computed
${\cal N}_{\alpha}$ values for $\alpha=0.75$ (boxed curve). We
have shown here for illustration, the plot for only
$\alpha=0.75$. It is observed that as $\alpha$ is increased from
0 to 1, the boxed curves (for different $\alpha$'s) shift
downward continuously indicating an $\alpha$ dependence in ${\cal
N}_{\alpha}$ for the extended matrix model of RNA. We investigate
this dependence in the following numerical analysis.

\begin{figure}
\includegraphics[width=4cm]{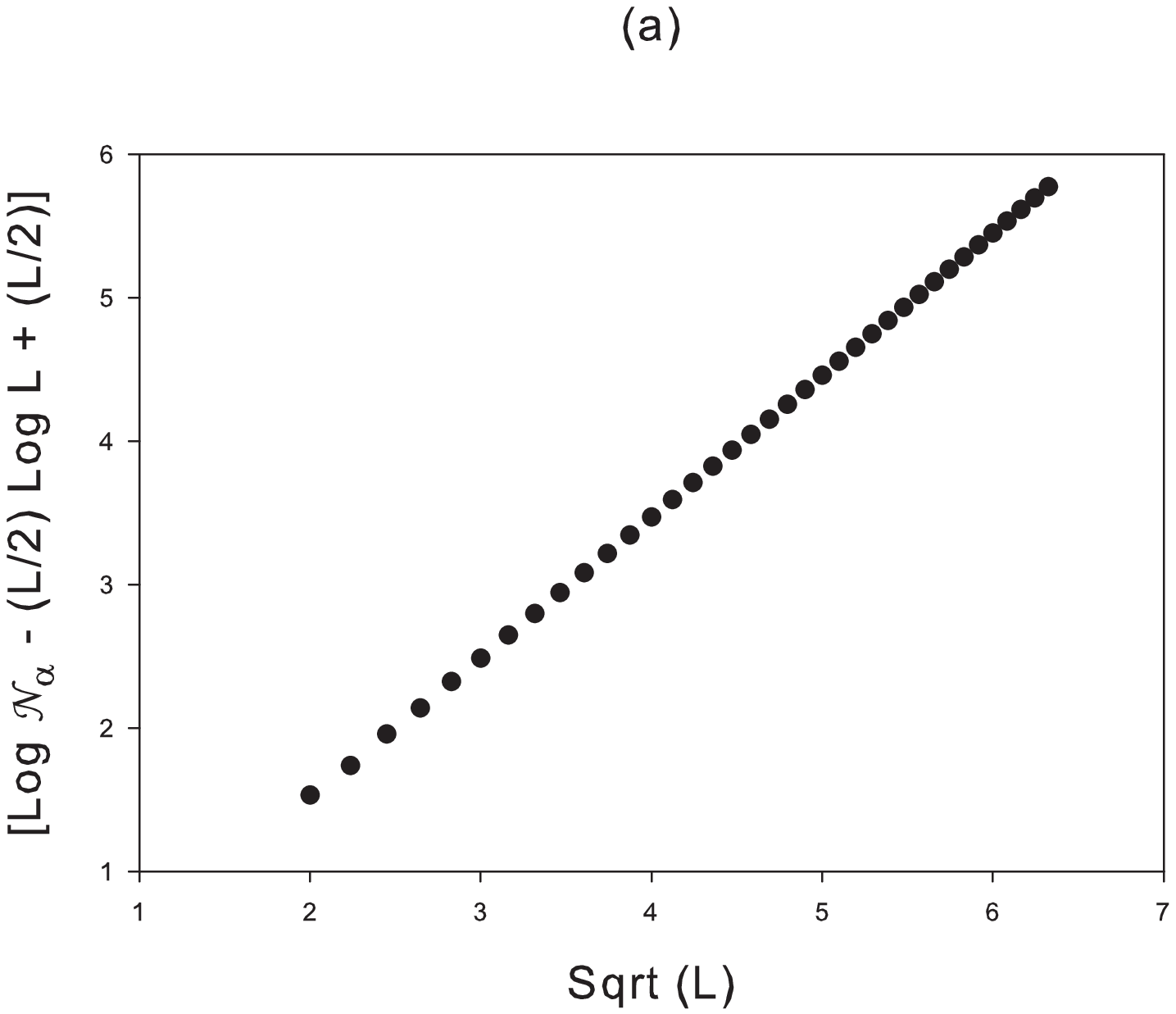}
\includegraphics[width=4cm]{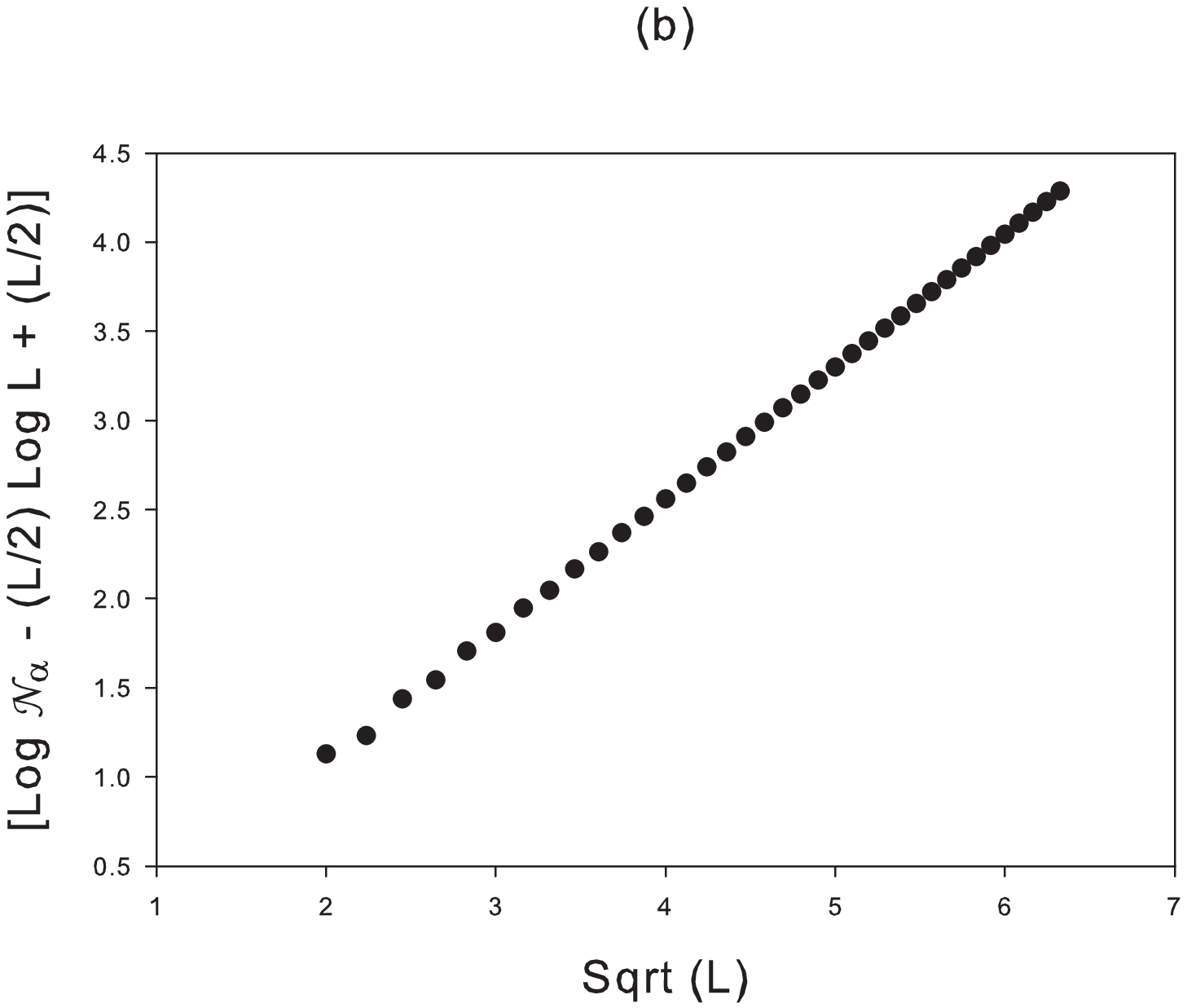}
\includegraphics[width=4cm]{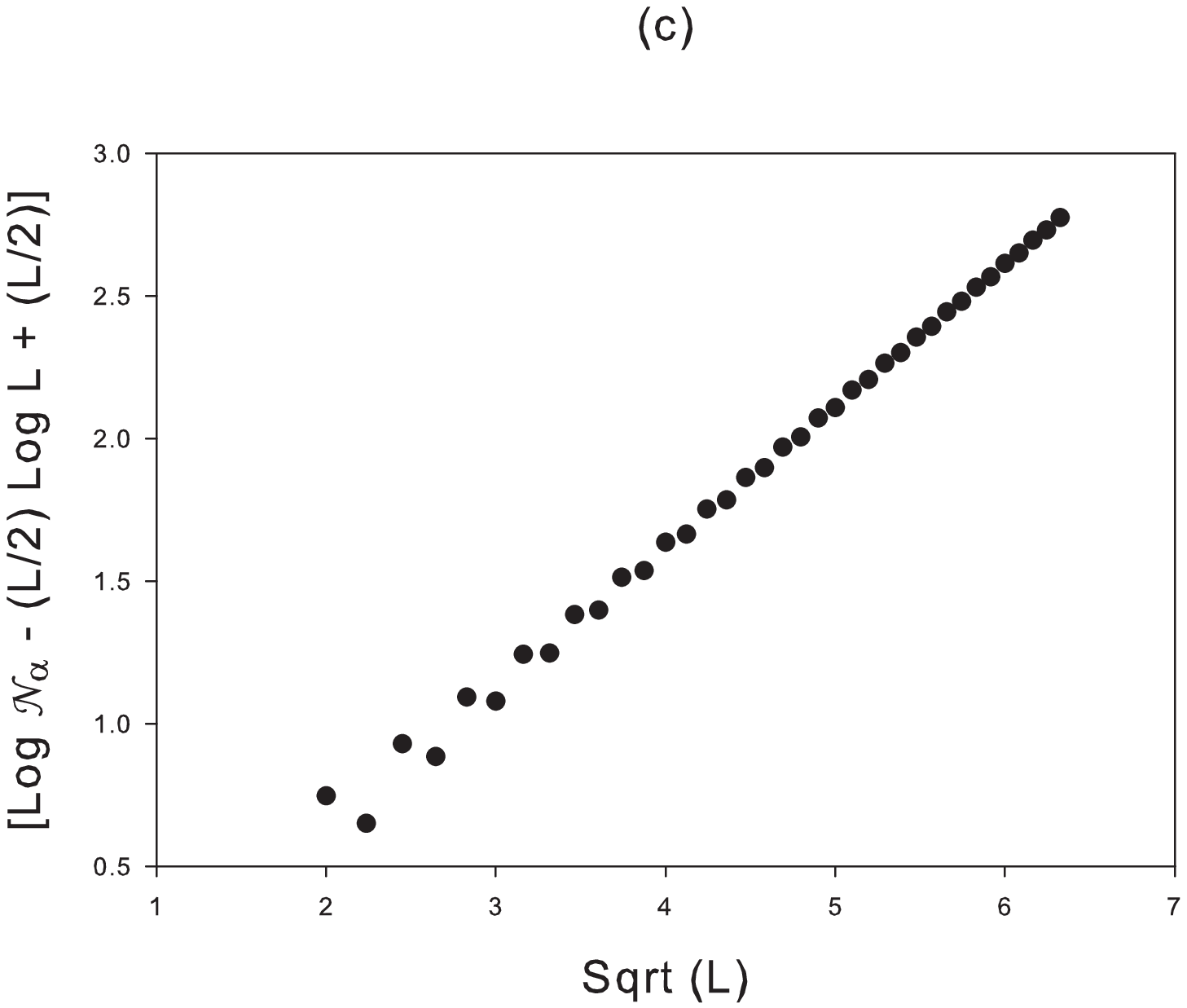}\\
\includegraphics[width=4cm]{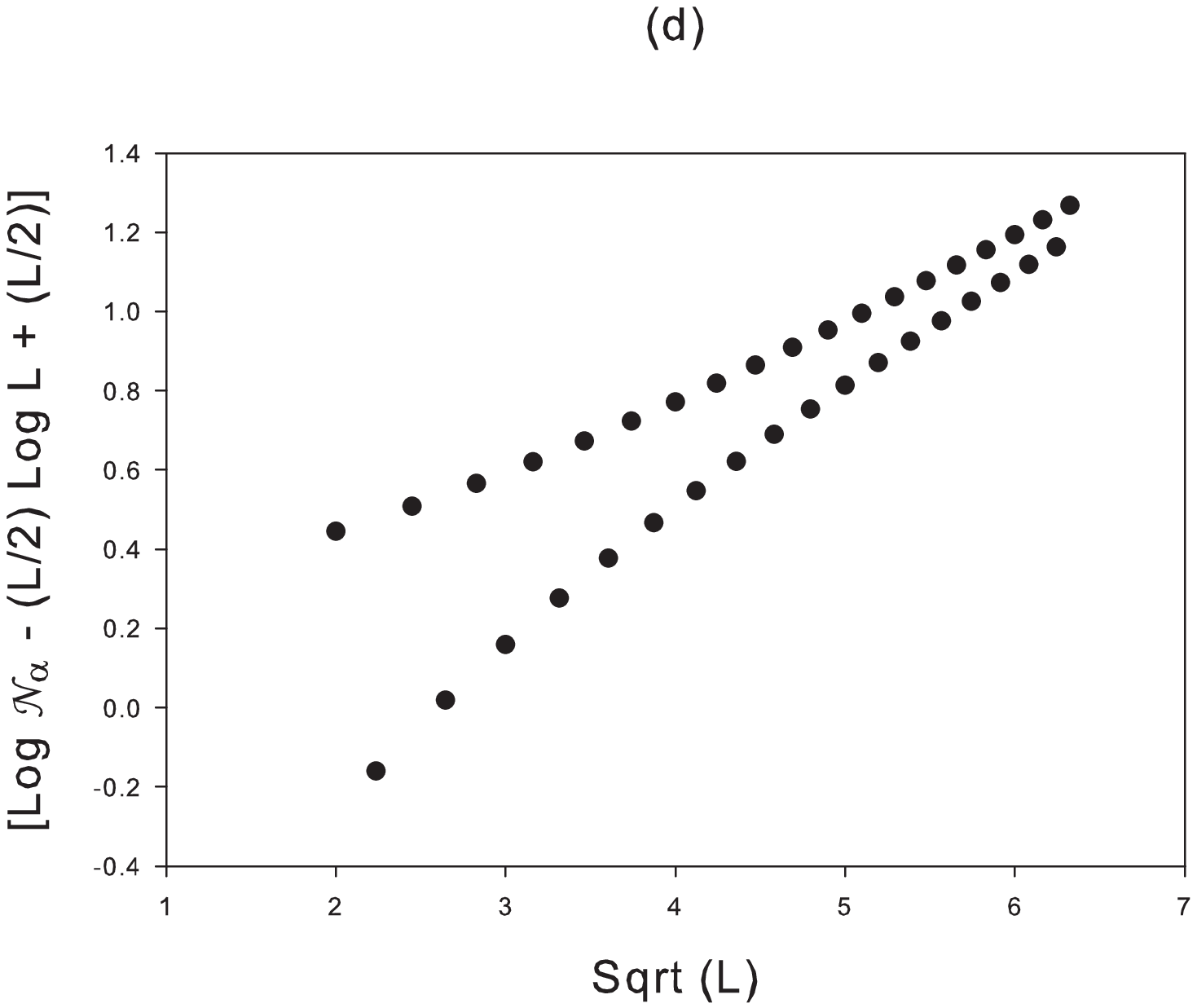}
\includegraphics[width=4cm]{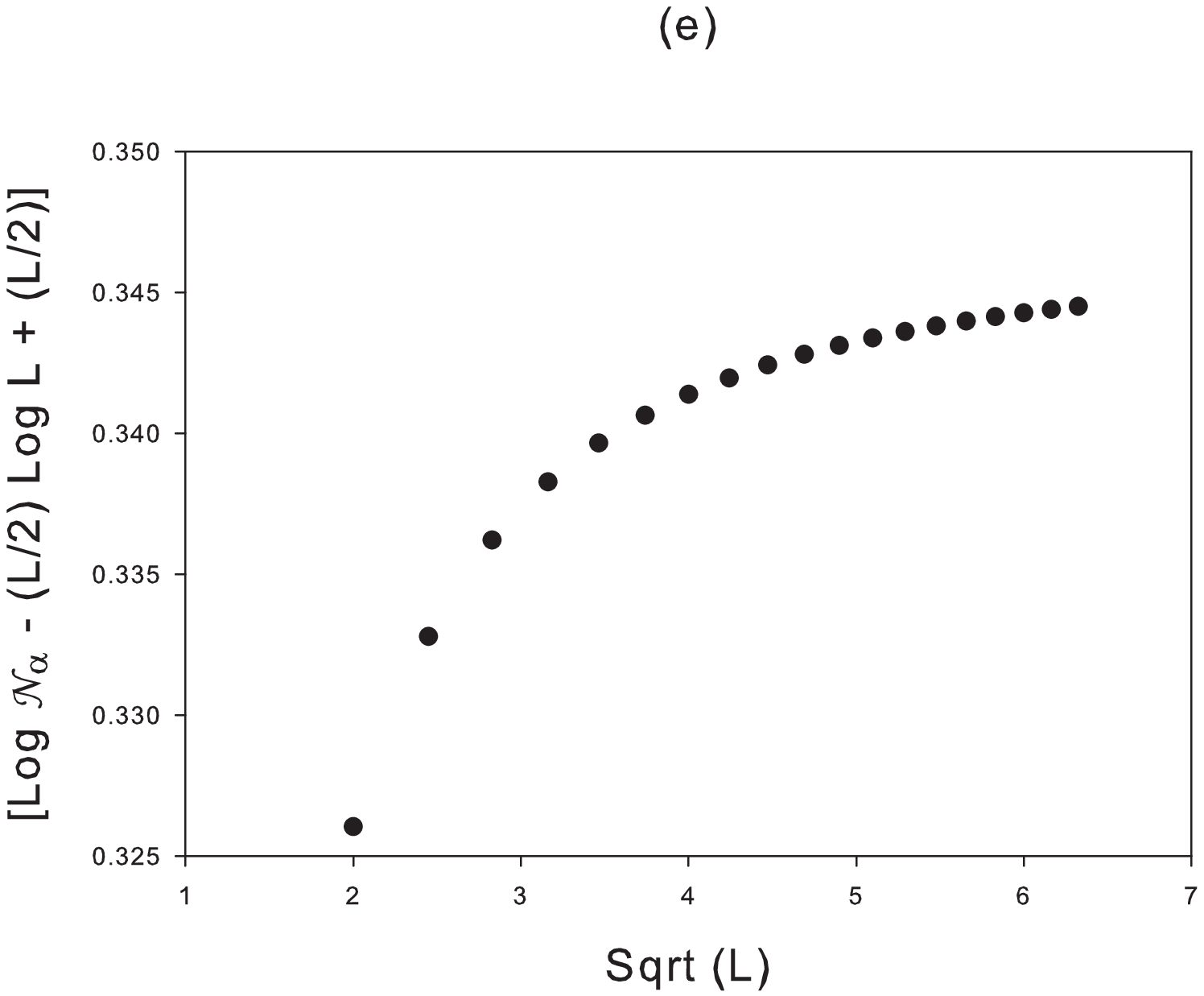}
\caption{$(Log {\cal N}_{\alpha}$ - $\frac{1}{2}$LLog L +
$\frac{L}{2})$ verses $\sqrt{L}$ plots for different values of
$\alpha$ along with their linearly fitted slopes: (a) $\alpha=0$
(slope=0.9818), (b) $\alpha=0.25$ (slope=0.7359), (c) $\alpha=0.5$
(slope=0.4926), (d) $\alpha=0.75$ (linear fit to the two curves
gives slope =0.3595) and (e) $\alpha=1$ (the plot is not linear).}
\label{fig.2}
\end{figure}

\begin{table}
\caption{Table lists slopes of the linearly fitted plots for
different values of $\alpha$ before and after the multiplication of
$(1-\alpha)$ with the $\sqrt{L}$ term of (i) $(Log {\cal
N}_{\alpha}$ + $\frac{L}{2}$ - $\sqrt{L}$) verses LLogL (Slope 1),
(ii) $[Log {\cal N}_{\alpha} + \frac{L}{2} - (1 - \alpha)\sqrt{L}$]
verses LLogL [Slope 1(a)], (iii) $(Log {\cal N}_{\alpha}$ -
$\sqrt{L}$ - $\frac{1}{2}$LLogL) verses L (Slope 2) and (iv) $[Log
{\cal N}_{\alpha} - (1 - \alpha)\sqrt{L} - \frac{1}{2} LLogL$]
verses L [Slope 2(a)].} \label{tab.1}
\vspace{.2in}
\begin{tabular}{lllll}
\hline \hline
$\alpha$ & Slope 1 & Slope 1(a) & Slope 2 & Slope 2(a)\\
\hline
0 & $0.499$ & $0.499$ & $-0.5026$ & $-0.5026$\\
\hline
0.25 & $0.4885$ & $0.499$ & $-0.5353$ & $-0.5022$\\
\hline
0.5 & $0.4767$ & $0.4987$ & $-0.5683$ & $-0.5027$\\
\hline
0.75 & $0.4624$ & $0.4981$ & $-0.6025$ & $-0.5060$\\
\hline
1 & $0.4556$ & $0.5003$ & $-0.6331$ & $-0.4992$\\
\hline \hline
\end{tabular}
\end{table}

Taking Log of $\cal N$ we get: $Log {\cal N} \sim \frac{L}{2} Log L
-\frac{L}{2} + \sqrt{L} - \frac{1}{4} - Log \sqrt{2}$. We are
interested in the large length L behaviour and we see that the
dependence of Log $\cal N$ on L is strongest in LLogL. We linearly
fit the plots (i) ($Log {\cal N}_{\alpha} - \sqrt{L} + \frac{L}{2}$)
verses LLogL (Slope 1, table~\ref{tab.1}), (ii) ($Log {\cal
N}_{\alpha} - \sqrt{L} - \frac{1}{2} LLogL$) verses L (Slope 2,
table~\ref{tab.1}) and (iii) ($Log {\cal N}_{\alpha} + \frac{L}{2} -
\frac{1}{2} LLogL$) verses $\sqrt{L}$ (fig.~\ref{fig.2}) for
different $\alpha$ and find their slopes. We find that there is a
continuous decrease in the slopes as $\alpha$ goes from 0 to 1 in
the linearly fitted plots of (i) and (ii) (Slope 1 and Slope 2
respectively of table~\ref{tab.1}), strongly suggesting a dependence
of ${\cal N}_{\alpha}$ on $\alpha$. In the fitted plots of (iii) we
observe a remarkable behaviour for $\alpha=0.75$ and $\alpha=1$
plots [fig.~\ref{fig.2}(d) and fig.~\ref{fig.2}(e)]. In the
$\alpha=0.75$ plot [fig.~\ref{fig.2}(d)], the points for odd and
even lengths separate out into two very distinct curves and for the
$\alpha=1$ plot [fig.~\ref{fig.2}(e)], the odd lengths vanish
completely leaving only the even length points in the figure. This
indicates that ($Log {\cal N}_{\alpha} + \frac{L}{2} - \frac{1}{2}
LLogL$) verses $\sqrt{L}$ is very sensitive to changes in $\alpha$.

We try a factor of $(1-\alpha)$ with the $\sqrt{L}$ term in the
exponent of the $\cal N$ expression and then fit the plots: (i)
$[Log {\cal N}_{\alpha} + \frac{L}{2} - (1 - \alpha)\sqrt{L}]$
verses LLogL [Slope 1(a), table~\ref{tab.1}] and (ii) $[Log {\cal
N}_{\alpha} - (1 - \alpha)\sqrt{L} - \frac{1}{2} LLogL]$ verses L
[Slope 2(a), table~\ref{tab.1}] for different values of $\alpha$. We
observe that now all the slopes are nearly the same and equal to
+$\frac{1}{2}$ and -$\frac{1}{2}$ for (i) and (ii) respectively.
This proves that the factor of $(1-\alpha)$ with the $\sqrt{L}$ term
in the exponent of $\cal N$ is the correct choice. We can therefore
write the new asymptotic expression of the total number of diagrams
at a fixed length L and $\alpha$ but independent of genus g, ${\cal
N}^\prime_{\alpha}$ for the extended matrix model as

\begin{equation}
\label{eq.4} {\cal N}^\prime_{\alpha} = L^{\frac{L}{2}}
exp^{\left[-\frac{L}{2} + (1 - \alpha) \sqrt{L} -
\frac{1}{4}\right]}/\sqrt{2}.
\end{equation}

We see from eq.~(\ref{eq.4}) that the total number of structures
for the extended matrix model changes considerably for example,
when $\alpha=1$ the $\sqrt{L}$ term vanishes from the exponent.
We repeat the exercise as before and plot the new asymptotic
formula ${\cal N}^\prime_{\alpha}$ for the extended matrix model
of RNA given by eq.~(\ref{eq.4}) [fig.~\ref{fig.1}($a^\prime$),
red dotted curve] with the numerically obtained ${\cal
N}_{\alpha}$ values for different $\alpha$'s (represented by
boxed curve, shown here for only $\alpha=0.75$). The plot for the
new asymptotic formula coincides with the numerical data ${\cal
N}_{\alpha}$ confirming the new formula.

\begin{figure*}
\includegraphics[height=14cm]{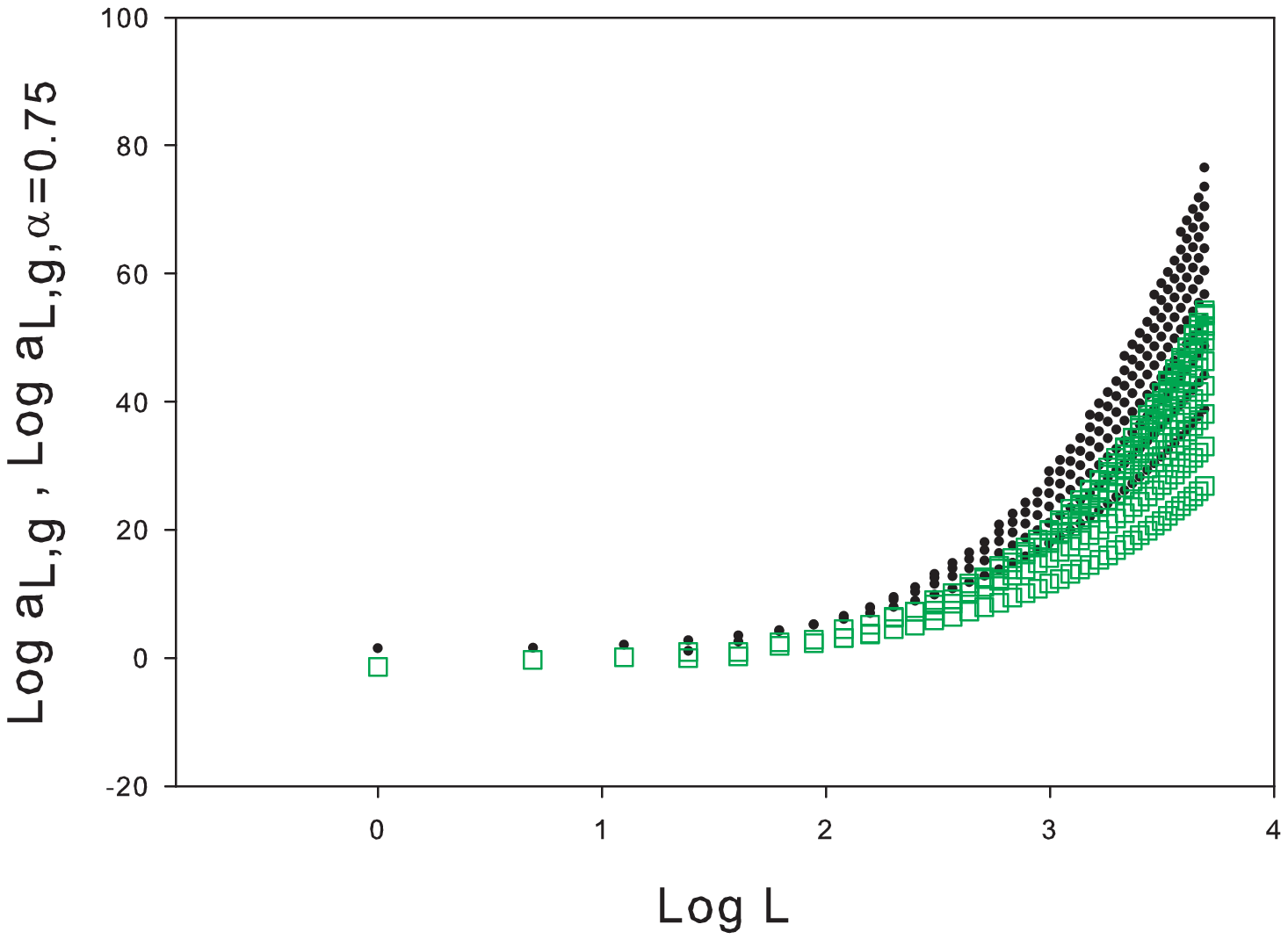}
\caption{The asymptotic formula for $a_{L, g}$ in \cite{18} (black
dotted curve) is plotted with the numerical $a_{L, g, \alpha}$
values (green boxed curve) for different lengths L for
$\alpha=0.75$.
\newline Note: The figure plots $a_{L, g, \alpha}$'s for all
genii corresponding to a particular length L of the polymer chain.
The lowest curve (black dotted or green boxed) corresponds to genus
$g=0$ for all the lengths (0 to 40) and the successive curves in the
upward direction correspond to next higher genii with the maximum
genus given by $g_{max}=L/4$.}
\label{fig.3}
\end{figure*}

\subsubsection{Asymptotics for $a_{L, g, \alpha}$}

The plot (fig.~\ref{fig.3}) of the asymptotic formula for $a_{L, g}$
(black dotted curve) with the numerically calculated $a_{L, g,
\alpha}$ values (green boxed curve) for different $\alpha$'s (shown
for $\alpha=0.75$) clearly indicates that the asymptotic formula of
the model in \cite{18} needs to be changed to give the asymptotic
behaviour of the extended matrix model of RNA folding \cite{16}. The
curves for different $\alpha$'s (shown here for only $\alpha=0.75$,
fig.~\ref{fig.3}) move further and further away from the asymptotic
expression curve \cite{18} as $\alpha$ goes from 0 to 1. This
behaviour is studied and the correct asymptotic expression
$a^\prime_{L, g, \alpha}$ for the extended matrix model is found.

We start with the asymptotic expression of $a_{L, g}= k_{g} 3^{L}
L^{(3g - \frac{3}{2})}$, where $k_{g} = \frac{1}{3^{(4g -
\frac{3}{2})} 2^{2g +1} g! \sqrt{\Pi}}$. Taking Log on both of the
sides and fixing $g=1$ (for simplicity) we get $Log(a_{L, g=1}) \sim
Log\frac{1}{3^{(\frac{5}{2})} 2^{(3)} \sqrt{\Pi}} + L Log 3 +
\frac{3}{2} Log L$. In Log ($a_{L, g=1}$), L dependence is present
in the form of L and LogL. We are interested in the large L
behaviour so we first look for the dominant L dependence. The linear
fits to the plots of $Log (a_{L, g=1, \alpha})$ verses L in
table~\ref{tab.2} (Slope 1) shows that the slopes of the numerical
$a_{L, g, \alpha}$ curves for different $\alpha$'s are not the same
and not equal to the slope of the $a_{L, g}$ asymptotic curve (slope
should be Log 3 for a plot between $Log(a_{L, g=1, \alpha})$ and L
according to \cite{18}). This indicates an $\alpha$ dependence in
the factor 3 of the $3^{L}$ universal part of $a_{L, g}$ which we
represent by $x(\alpha)$ in table~\ref{tab.2} [where $x(\alpha)=Log
(Slope 1)$]. We write the asymptotic formula by replacing 3 with
$x(\alpha)$. The expression for $a^\prime_{L, g, \alpha}$ after
taking Log on both of the sides becomes $Log{a^\prime_{L, g,
\alpha}} \sim Log k_{g} + L Log[x(\alpha)] + \left[3g -
\frac{3}{2}\right] Log{L}$. To determine the form of x($\alpha$), we
plot $x(\alpha)$ verses $\alpha$ which is a straight line with
$slope=-1.133$ and $intercept=3.466$. In the same way as the
asymptotic expression for $a_{L, g}$ in \cite{18} had the universal
term $3^{L}$, we find $x(\alpha)^{L}$ to be $x(\alpha)^{L} = (-
\alpha + 3)^{L}$ for all $\alpha$. We therefore have
$Log{a^\prime_{L, g, \alpha}} \sim Log k_{g} + L Log (3 - \alpha) +
\left[3g - \frac{3}{2}\right] Log L$. The universal $3^{L}$ part in
the $a_{L, g}$ \cite{18} has been modified to $(3 - \alpha)^{L}$ for
the extended matrix model \cite{16}. The asymptotic formula gets
modified to $a^\prime_{L, g, \alpha}$ $\sim$ $k_{g} (3 - \alpha)^{L}
L^{(3g - \frac{3}{2})}$.

\begin{table}
\caption{Table lists the measures of slopes for different values of
$\alpha$ obtained from the linear fits to the plots between L and
Log $a_{L, g=1, \alpha}$ (Slope 1), the $x(\alpha)$ values for each
$\alpha$ and slopes from the linear fit of plots between Log L and
[Log $a_{L, g=1, \alpha} - L Log (3 - \alpha)]$ for each $\alpha$
(Slope 2).}
\label{tab.2}
\begin{center}
\begin{tabular}{llll}
\hline
\hline
$\alpha$ & Slope 1 & $x(\alpha)$ & Slope 2 \\
\hline
0 & 1.198 & 3.313 & 1.646\\
\hline
0.25 & 1.109 & 3.03 & 1.639\\
\hline
0.5 & 1.012 & 2.75 & 1.633\\
\hline
0.75 & 0.9065 & 2.476 & 1.623\\
\hline
1 & 0.7891 & 2.2 & 1.655\\
\hline
Analytical & 1.24 & 3.4556 & 1.495\\
\hline
\hline
\end{tabular}
\end{center}
\end{table}

\begin{figure}
\includegraphics[width=4cm]{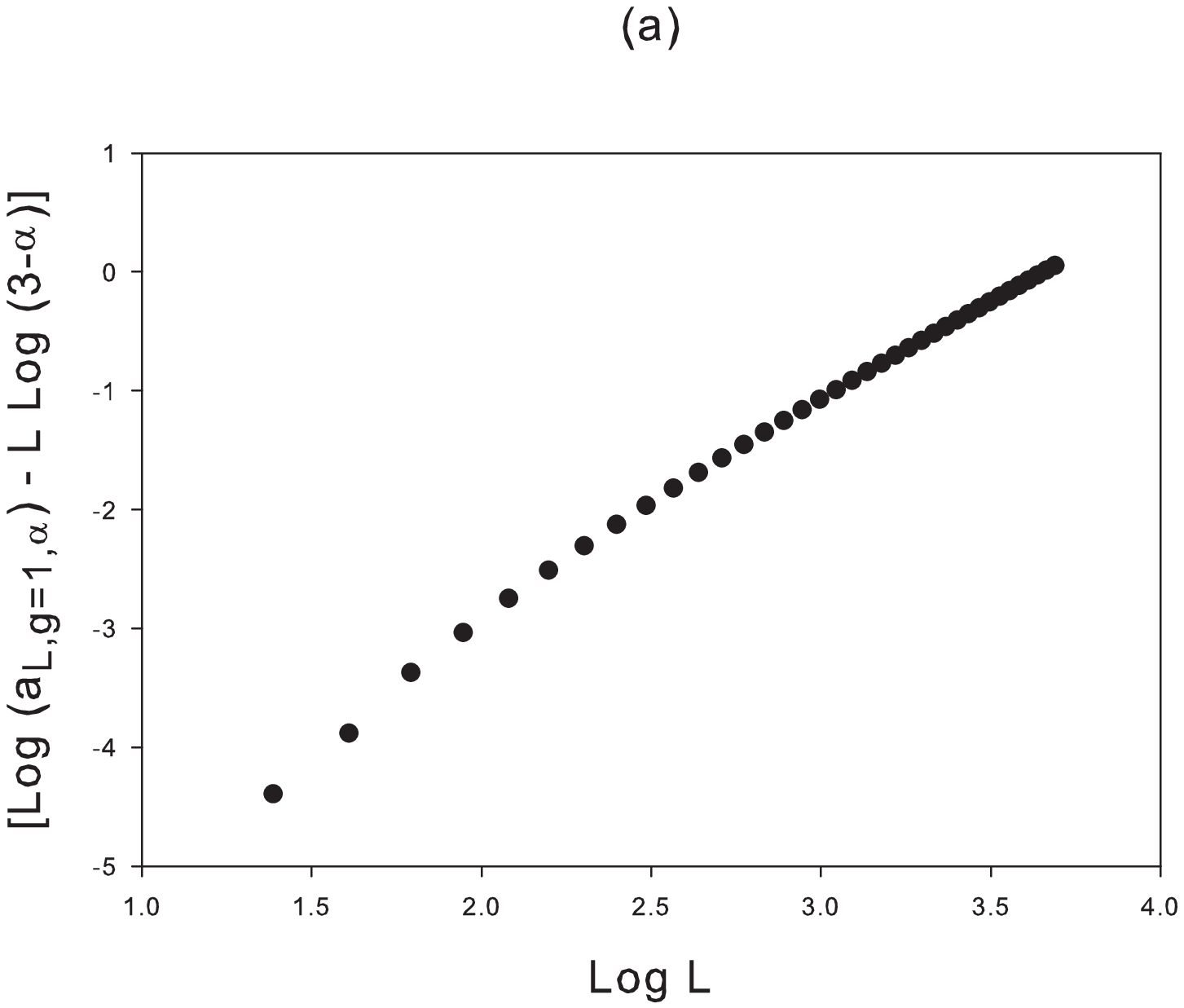}
\includegraphics[width=4cm]{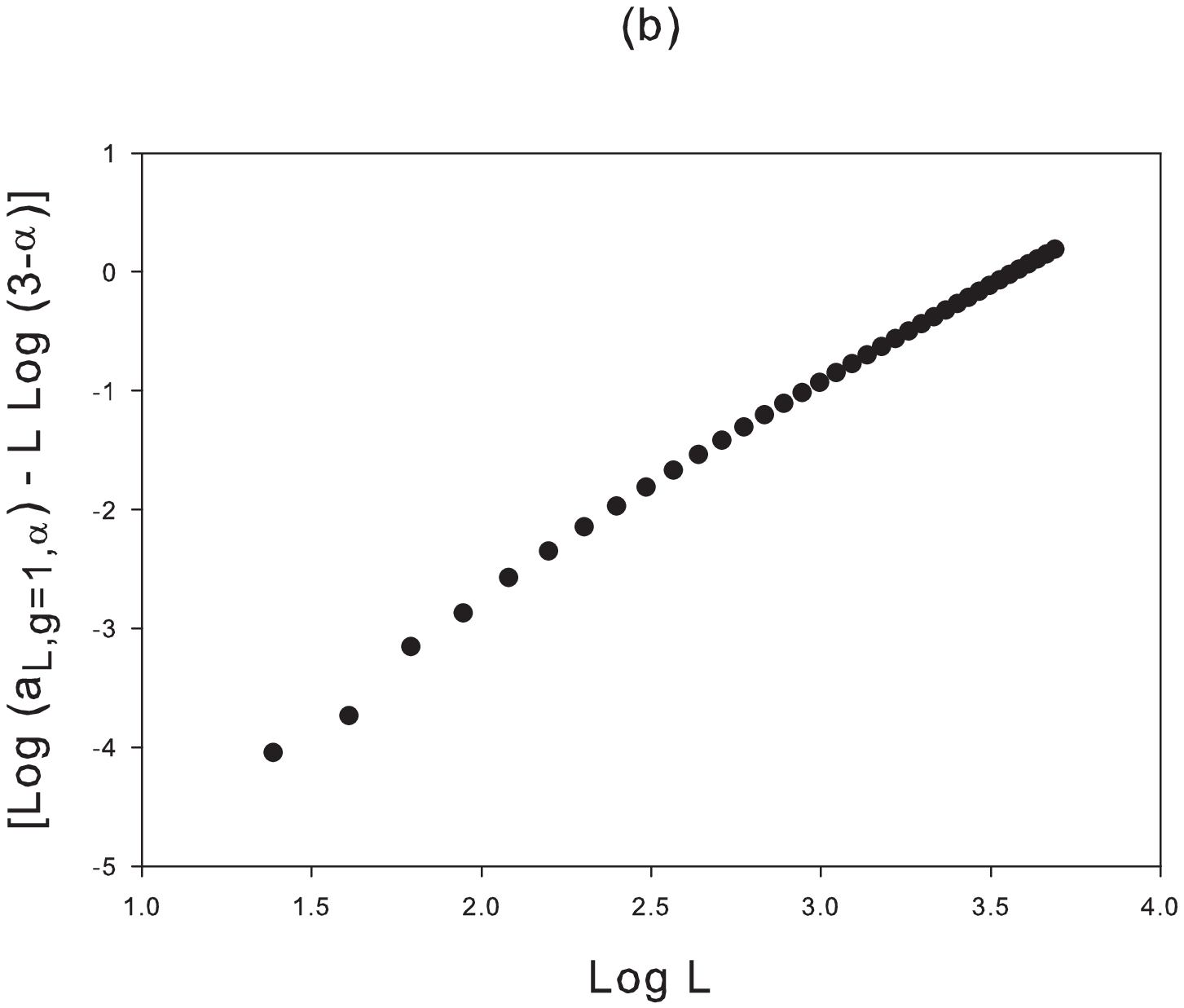}
\includegraphics[width=4cm]{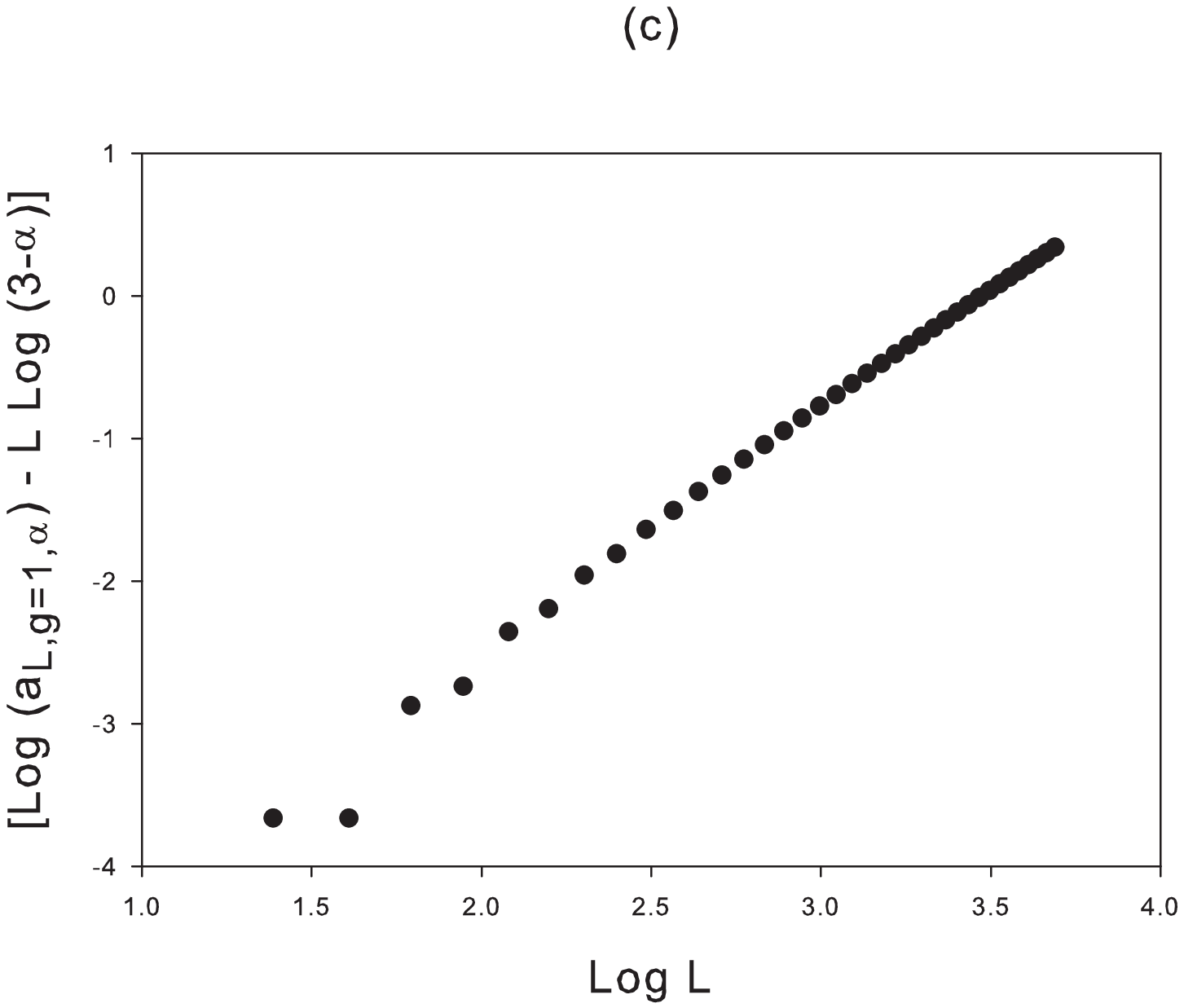}\\
\includegraphics[width=4cm]{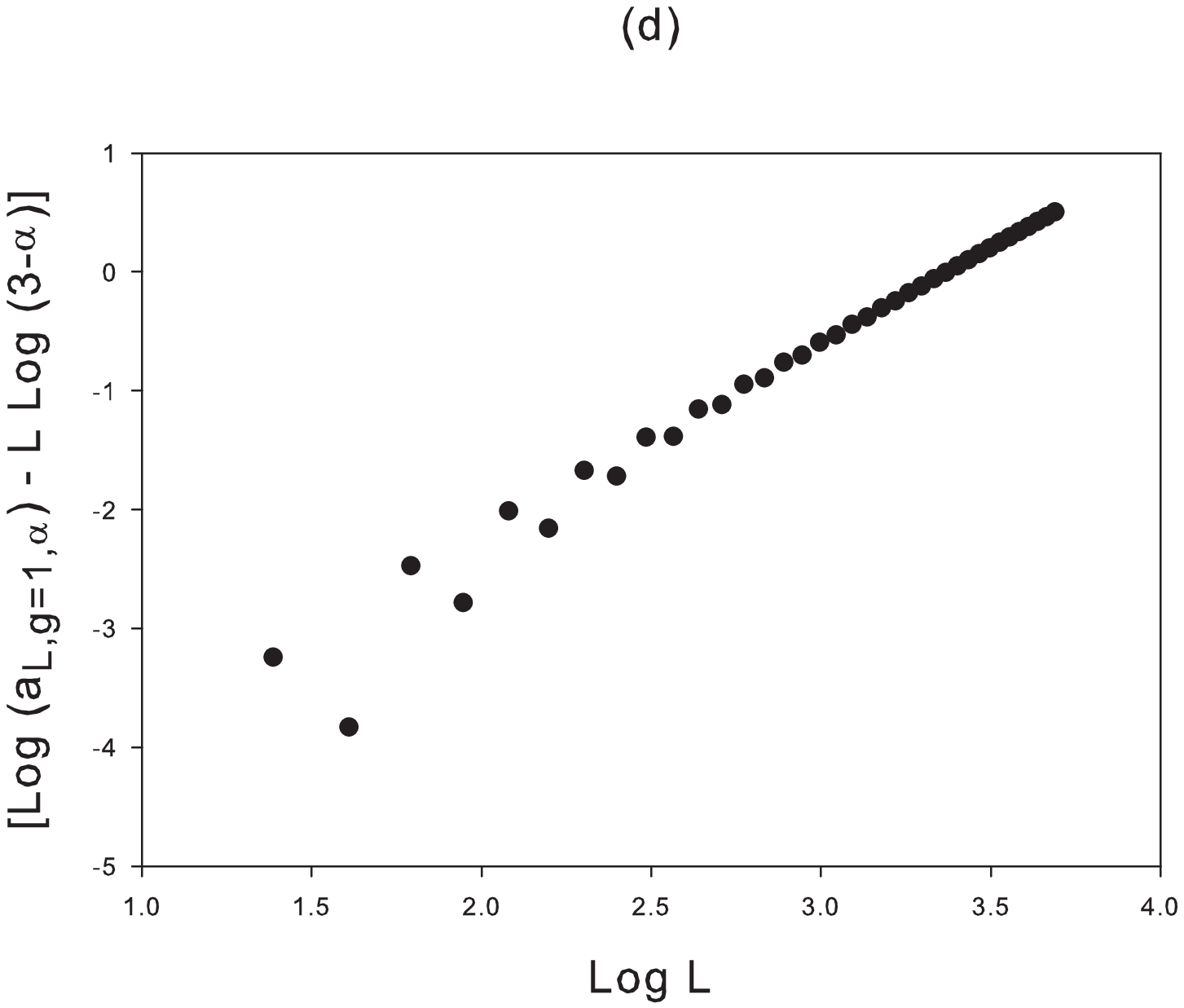}
\includegraphics[width=4cm]{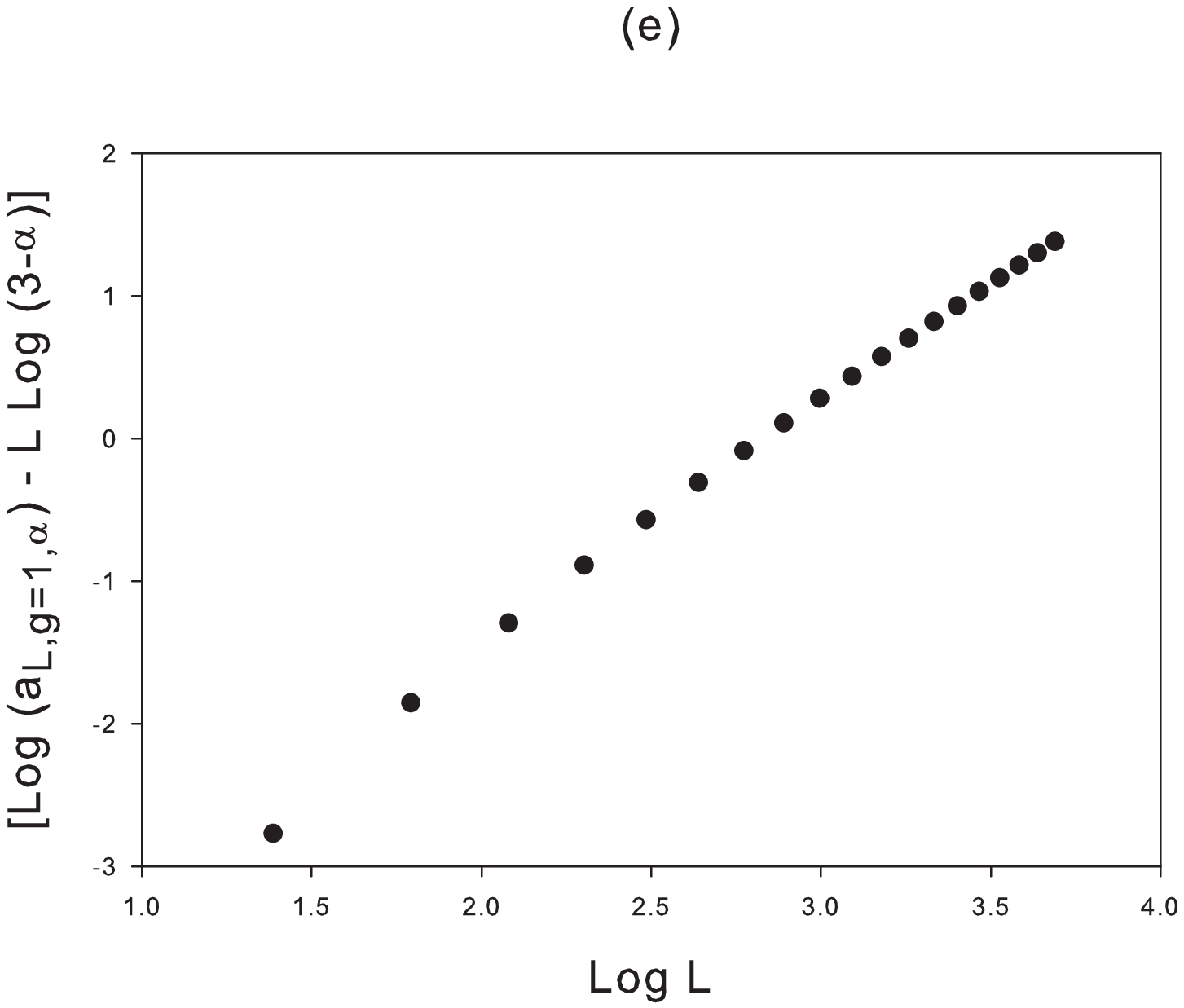}
\caption{$[Log a_{L,g=1, \alpha} - L Log(3-\alpha)]$ verses Log L
plots for different values of $\alpha$, (a) $\alpha=0$, (b)
$\alpha=0.25$, (c) $\alpha=0.5$, (d) $\alpha=0.75$ and (e)
$\alpha=1$. The slopes for these values of $\alpha$ are listed in
Table 2 (Slope 2).} \label{fig.4}
\end{figure}

Analyzing the Log(L) dependence now, we assume that there exists an
$\alpha$ dependence in the exponent of L which we represent by
$f(\alpha)$. We can therefore write from the modified equation after
taking Log on both the sides and substituting $g=1$,
$Log(a^\prime_{L, g=1, \alpha}) \sim Log\frac{1}{3^{(\frac{5}{2})}
2^{(3)} \sqrt{\Pi}} + L Log (3 - \alpha)+
\frac{3}{2}\left[f(\alpha)\right] Log L$. Linear fitted plots of
[$Log (a_{L, g=1, \alpha})-L Log (3 - \alpha)$] verses Log (L) for
different $\alpha$ values is shown in fig.~\ref{fig.4}. The figure
shows a continuous separation of data points belonging to the even
and odd lengths as $\alpha$ is increased from 0 to 1. There are two
distinct lines at small lengths L which merge into a single line at
higher lengths L. For $\alpha=1$ the points for odd lengths vanish
completely from the plot. The slopes [table~\ref{tab.2}, Slope 2]
show that the difference between analytical and numerical values for
different $\alpha$ is $\sim$ 0.01, which is small. The Log(L) term
therefore shows no significant $\alpha$ dependence. So we fix
$f(\alpha) = 1$. This gives the asymptotic formula of the number of
diagrams at a fixed length L, genus g and $\alpha$, $a^\prime_{L, g,
\alpha}$ for the extended matrix model of RNA as

\begin{equation}
\label{eq.5} a^\prime_{L, g, \alpha} \sim k_{g} (3 - \alpha)^{L}
L^{(3g - \frac{3}{2})}
\end{equation}

\begin{figure*}
\includegraphics[height=14cm]{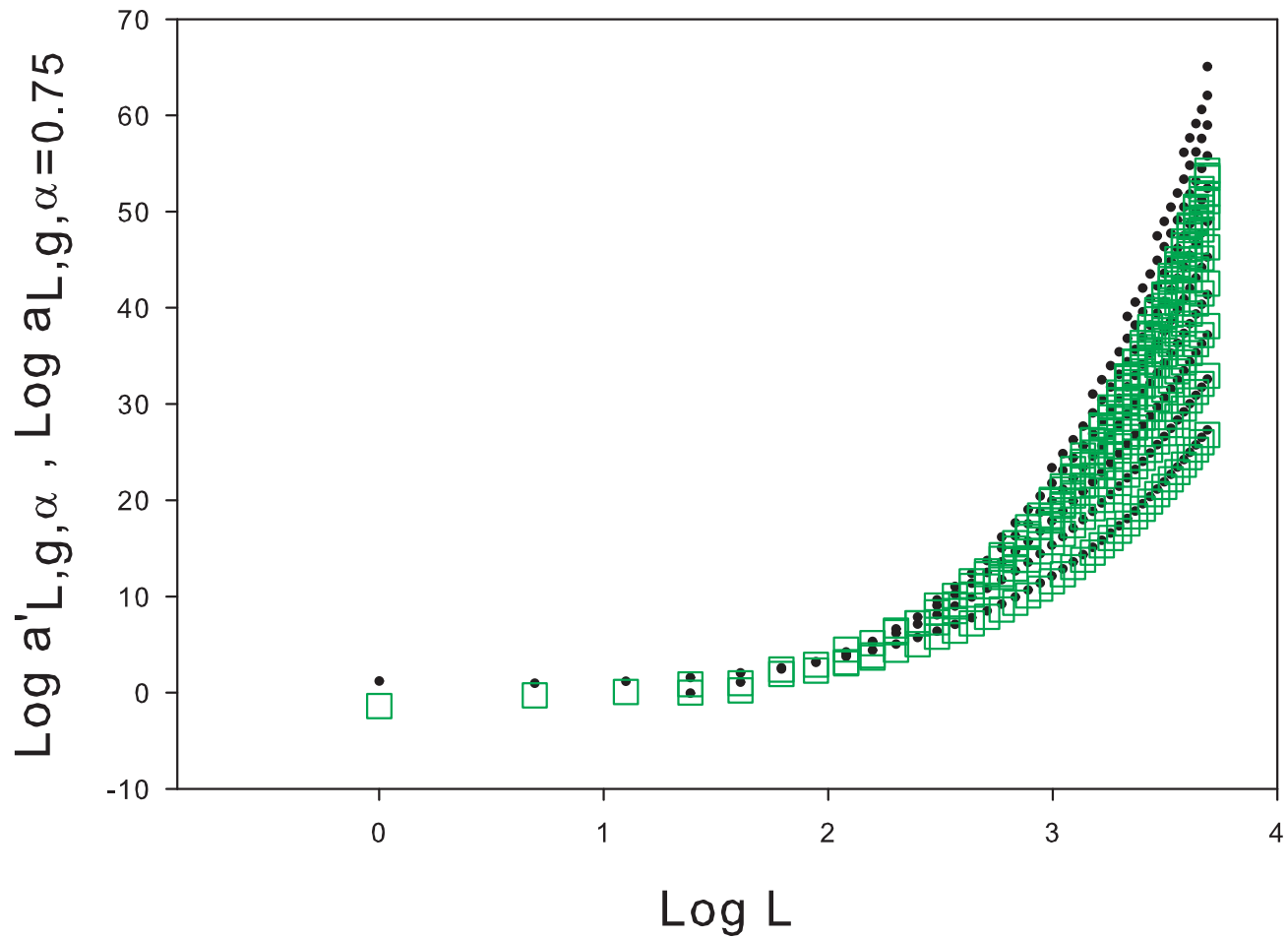}
\caption{The plot for the new asymptotic formula for $a^\prime_{L,
g, \alpha}$ (black dotted curve) for the extended matrix model of
RNA is shown with the numerically obtained $a_{L, g, \alpha}$ for
$\alpha=0.75$ (green boxed curve).}
\label{fig.5}
\end{figure*}

The asymptotic formula [eq.~(\ref{eq.5})] thus obtained is plotted
with the numerically found $a_{L, g, \alpha}$ values for different
$\alpha$'s [fig.~\ref{fig.5}, shown here for only $\alpha=0.75$] and
it is seen that the formula matches with the numerical results for
large L. To verify the final form of the formula, we substitute
different $\alpha$'s and $g=1$ in eq.~(\ref{eq.5}) and plot $[Log
a^\prime_{L, g=1, \alpha} - L Log (3 - \alpha)]$ verses LogL. The
slopes are found to be 1.495 in all the cases. This result will hold
for any genus g, though we have shown here the result for only $g =
1$. It is interesting to note here that the universality of
$a^\prime_{L, g, \alpha}$ for the extended matrix model changes from
$3^{L}$ in \cite{18} to $2^{L}$ when $\alpha=1$ (the completely
paired base region).

The asymptotic behaviour of $a_{L, g, \alpha, n}$ and ${\cal
N}_{\alpha, n}$ for the model with perturbation on n bases is the
same as for the model with perturbation on all the bases except
that $\alpha$ is replaced by $\frac{n\alpha}{L}$ [as is evident
from the expression of the exponential generating function $G(t,
N, \alpha)$ given by eq.~(\ref{eq.3}) with $\frac{n\alpha}{L}$ in
place of $\alpha$]. Thus we can write the asymptotic expressions
of the genus distribution functions for a perturbation acting on
n bases as

\begin{equation}
\label{eq.6} a^{\prime}_{L, g, \alpha, n} \sim k_{g} (3 -
\frac{n\alpha}{L})^{L} L^{(3g - \frac{3}{2})}
\end{equation}

and,

\begin{equation}
\label{eq.7} {\cal N}^{\prime}_{\alpha, n} = L^{\frac{L}{2}}
exp^{\left[-\frac{L}{2} + (1 - \frac{n\alpha}{L}) \sqrt{L} -
\frac{1}{4}\right]}/\sqrt{2}.
\end{equation}

The asymptotics for the extended matrix models therefore show marked
changes in the presence of the perturbation in the universal term of
$a_{L, g}$ and in the total number of structures ${\cal N}$ of the
model in \cite{18}.

\section{CONCLUSIONS}

In this work, we develop on the footsteps of the extended matrix
model of RNA folding proposed in \cite{16}, the effect of an
external perturbation on only one nucleotide in the polymer chain of
length L. We argue that $\alpha$ in the exponential generating
function of the partition function of model in \cite{16} will be
replaced by $\frac{\alpha}{L}$ if perturbation acts on only one
nucleotide in the chain. Further, we generalize this result to a
finite number $n \leq L$ of perturbations on the nucleotides of the
chain, where $\alpha$ in the exponential generating function of the
partition function gets replaced by $n\alpha/L$, eq.~(\ref{eq.3}).

Next, we find numerically the asymptotic behaviour of the genus
distribution functions for the extended matrix model of RNA folding
in \cite{16} and the n-NP model. We find from the numerical analysis
that the universality of $a_{L, g}$, $3^{L}$ found in \cite{18},
changes to $(3-\alpha)^{L}$ when the perturbation acts on all the
bases in the polymer chain [which becomes
$(3-\frac{n\alpha}{L})^{L}$ when the perturbation is on n bases].
The power law term $L^{3g-\frac{3}{2}}$ of $a_{L, g}$ \cite{18}
remains the same for the asymptotic formula of $a^\prime_{L, g,
\alpha}$ in the extended matrix models with perturbation on all the
bases \cite{16} and on n bases. The total number of diagrams ${\cal
N}$ also changes from its form in \cite{18} to ${\cal
N}^\prime_{\alpha} = L^{\frac{L}{2}} exp^{\left[-\frac{L}{2} + (1 -
\alpha) \sqrt{L} - \frac{1}{4}\right]}/\sqrt{2}$ with the term
$exp^{\sqrt{L}}$ in \cite{18} changing to $exp^{(1-\alpha)\sqrt{L}}$
for the matrix model with perturbation on all the bases [which
becomes $exp^{(1-\frac{n\alpha}{L})\sqrt{L}}$ when the perturbation
is on n bases]. The most striking change found in the universality
of $a^\prime_{L, g, \alpha}$ is when $\alpha$ takes the value 1 (and
$n=L$) as the universality goes from $3^{L}$ to $2^{L}$ and in the
$(1-\alpha)\sqrt{L}$ term in the exponent of ${\cal
N}^\prime_{\alpha}$ which goes to zero when $\alpha=1$ and $n=L$. It
is shown in fig.~\ref{fig.2} and fig.~\ref{fig.4} that as $\alpha$
is increased from 0 to 1 in steps of 0.25, the points corresponding
to even and odd lengths of the chain start splitting up into two
different curves at small lengths, but converge into a single linear
curve as the length is increased. Note that at small lengths, this
difference is most pronounced for $\alpha=0.75$, for both ${\cal
N}_{\alpha}$ and $a_{L, g, \alpha}$. The $\alpha=1$ plots of ${\cal
N}_{\alpha}$ and $a_{L, g, \alpha}$ [fig.~\ref{fig.2}(e) and
fig.~\ref{fig.4}(e) respectively] show the absence of odd length
data points. It is interesting to note that the genus distributions
show different behaviour at small and large lengths (analysis has
been done for $L=40$). The large L (asymptotic) behaviour of the
distribution functions [eq.~(\ref{eq.4}), (5), (6) and (7)] found
for the RNA matrix model with external perturbation show prominent
changes.

We have studied the effect of an external perturbation on the
RNA matrix model. In order to compare the results of the matrix
model of RNA folding with external perturbations (discussed here and
in \cite{16}) with experiments (where the perturbations may be
due to the constant forces discussed in the introduction or due to
natural processes like transcription and translation taking
place inside a living cell), a more detailed study will be
undertaken.

\section*{ACKNOWLEDGEMENTS}

We would like to thank Profs. H. Orland and G. Vernizzi for very
valuable and encouraging e-discussions. The work was financially
supported by CSIR Project No. $03(1019)/05/EMR-II$.

\end{document}